\documentclass[aps,graphicx,prl,twocolumn,superscriptaddress]{revtex4}
\usepackage[dvips]{graphics}
\usepackage[dvips]{graphicx}

\def\eps{\varepsilon}
\def\be{\begin{equation}}
\def\ee{\end{equation}}
\def\ben{$$}
\def\een{$$}
\def\beq{\begin{eqnarray}}
\def\eeq{\end{eqnarray}}
\begin{document}
\title{Orbital ordering in manganites in the band approach.}
\author{D.V.~Efremov}
\affiliation{%
Institut fur Theoretische Physik, TU Dresden
, 01062 Dresden
}
\affiliation{%
Laboratory of Solid State Physics, Material Science Center,
University of Groningen, Nijenborgh 4, 9747 AG Groningen, The
Netherlands }
\author{D.I.~Khomskii}
\affiliation{%
University of Cologne, Zuelpicher Str. 77, 50937 Cologne, Germany
}
\affiliation{%
Laboratory of Solid State Physics, Material Science Center,
University of Groningen, Nijenborgh 4, 9747 AG Groningen, The
Netherlands }
\date{\today}
\begin{abstract}
We consider the orbital ordering in LaMnO$_3$ and similar systems,
proceeding from the band picture. We show that for the realistic
magnetic structure of A-type there exists a complete nesting
betweeen two $e_g$-bands. As a result there occurs an instability
towards an excitonic insulator-like state -- an
electron-hole pairing with the wave vector $Q=(\pi,\pi)$, which opens a gap in
the spectrum  and makes the system insulating. In the resulting
state there appeasr an orbital ordering -- orbital density wave
(ODW), the type of which coincides with those existing in
LaMnO$_3$.
\end{abstract}
\maketitle
\section{Introduction}

Manganite oxides R$_{1-x}$M$_{x}$MnO$_3$ (R=La, Pr, Nd,... and M =
Ca, Sr) attract  great interest both from experimental and
theoretical points of view. The phase diagram of these compounds is
very rich, and in large part of it there exist, besides magnetic, also an
orbital ordering  due to the Jahn-Teller nature of
Mn$^{+3}$ ions with configuration $t_{2g}^3e_g$. Usually this
ordering, especially at the end of the series RMnO$_3$
(x=0) is described in the picture of localized $d$-electrons \cite{vandenBrink}.
However it was recently shown
\cite{vandenBrink99a,vandenBrink99b,Efremov03} that the orbital
structure of doped materials can be rather successfully explained from
the opposite, band structure point of view -- at least for the
$e_g$-electrons. Experimentally manganites are not very far
from the metallic state. Although being insulating, undoped LaMnO$_3$  has
relatively small energy gap, by optical data of the order of 0.4 eV \cite{Tokura}.
Moreover the conductivity of RMnO$_3$  considerably increases above the  temperature of
the orbital ordering \cite{Zhou}.
Thus the question whether one can explain the main
properties of undoped RMnO$_3$ materials, in particular their
insulating character and the orbital ordering, proceeding from the band
structure point of view, has a great interest.

The first attempt in this direction was undertaken by Gor'kov and
Kresin \cite{Gorkov} who wrote down  the spectrum of $e_g$ - electrons on a cubic lattice in a general form
based only on  symmetry arguments.
However they did
not take into account  particular relations between hopping matrix elements
for the $e_g$-electrons, and did  not discuss the specific features of the
Fermi-surface, such as nesting, which play crucial role in our treatment

In the present paper, following the works
\cite{Gorkov,vandenBrink99a,Efremov03}, we consider the ground
state of LaMnO$_3$ from the band structure point of view. We show
that in the framework of the double-exchange picture there exist
perfect nesting between the "electron" and "hole" pockets of two
$e_g$- bands for the realistic magnetic structure of A-type
(ferromagnetic planes stacked antiferromagnetically), and also for
the hypothetical ferromagnetic state. As a result of the nesting
the system is unstable against electron-hole pairing of the
excitonic insulator type. Correspondingly the quasiparticle
spectrum develops an energy gap, and the system becomes
insulating. The pairing occurs with the wavevector $Q=(\pi , \pi
)$  forming  two sublattices which correspond to orbital ordering.
The type of this orbital ordering obtained theoretically coincides
with that observed in LaMnO$_3$ experimentally, although the
magnitude of it and of corresponding lattice distortions in this
weak coupling scheme is smaller than that existing in reality.
Thus we show that the type of the ground state, its insulating
character and the symmetry of orbital ordering is reproduced
correctly not only from the localized, but also from the itinerant
point of view.

We do not claim, of course, that the real LaMnO$_3$ is fully
described by the weak coupling approach: in real systems the
electron correlations are definitely  substantial. Nevertheless,
the actual material may  show the features typical for both the
limiting situations, and therefore we believe that the treatment
carried out in this paper, which proceeds from the band picture, may be quite
useful and instructive.

\section{The model}
 In our calculations  we use the
tight binding approach for the band structure of manganites,
where both double exchange and superexchange are incorporated.
 In the double-exchange framework with the strong on-site
Hund's rule coupling the motion of $e_g$-electrons is largely
determined by the underlying magnetic structure. In this scheme
the stability of various magnetic and orbital structures is
determined by the competition of the band energy of the
$e_g$-electrons, favoring  ferromagnetism, and superexchange
interaction $J$ between localized ($t_{2g}$) spins, favoring
antiferromagnetism.

The effective  hopping matrix elements $t_{ij}^{\alpha \beta}$
depend on the type of the orbitals $\alpha, \beta$ and relative
orientation of a pair $i,j$, so that $t_{i,j \parallel x}^{x^2-y^2
, \,  x^2-y^2 } = 3/4 t $, $t_{i,j \parallel z}^{z^2 , \, z^2} =
t$, $t_{ij \parallel x}^{z^2, \,z^2} = 1/4 t$, etc
\cite{orbitalnote}. Besides this, there is the usual dependence of
the hopping on the angle $\theta$ between neighboring spins:
$\tilde{t}_{i,j}^{\alpha, \beta} = t_{ij}^{\alpha, \beta} \cos
\theta_{ij}/2$. The antiferromagnetic superexchange energy per
bond is $J \cos \theta_{ij}$.

Thus for this model with the infinite Hund energy we come to the
double exchange Hamiltonian:
\begin{equation} \label{eq:Hamiltonian}
 H = -\sum \tilde{t}^{\alpha \beta}_{ij} c^{\dagger}_{\alpha;
\sigma i}c^{}_{\beta; \sigma j} + J \sum \mathbf{S}_{i}
\mathbf{S}_{j} ,
\end{equation}
to which we will add later the electron-electron interaction.
The localized ($t_{2g}$) spins are treated classically. The last term in the Hamiltonian is
repulsive electron-electron on-site interaction.

\begin{figure}
 \includegraphics[width=0.4\columnwidth]{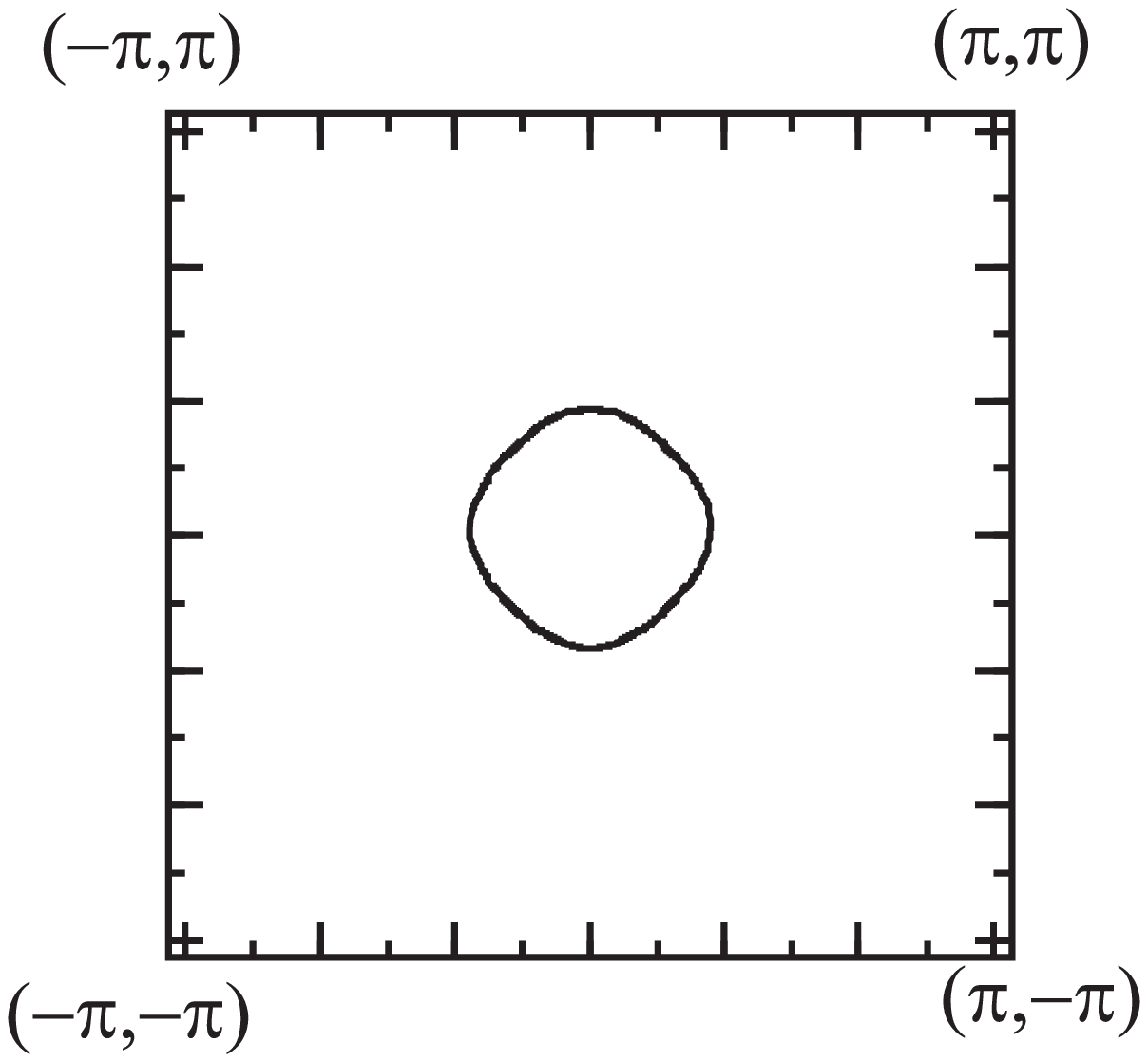}
  \includegraphics[width=0.4\columnwidth]{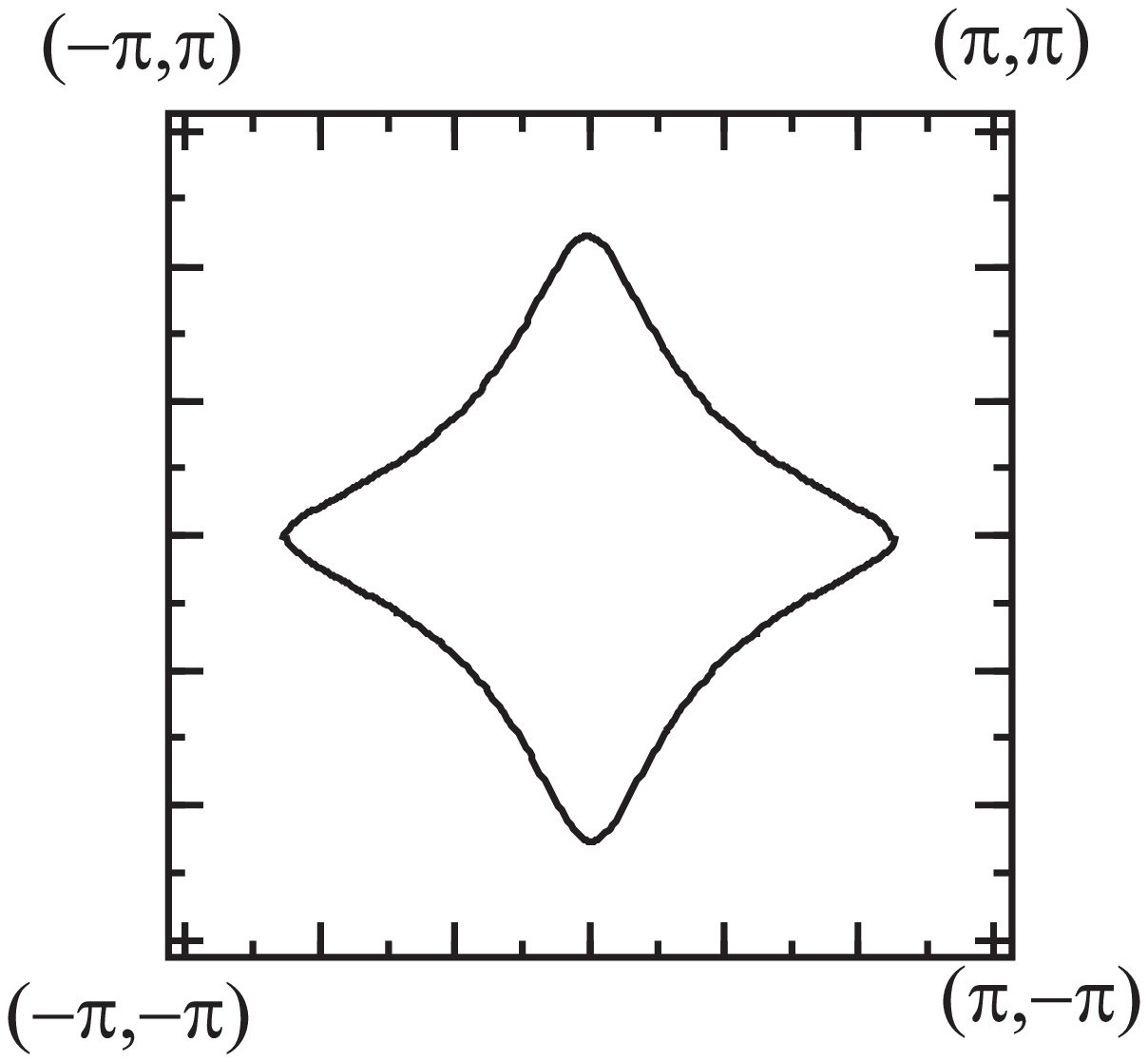}\\
\vspace{9pt}
  \hbox{\hspace{0.1\columnwidth} (a) \hspace{0.4\columnwidth} (b)}
  \vspace{9pt}
\includegraphics[width=0.4\columnwidth]{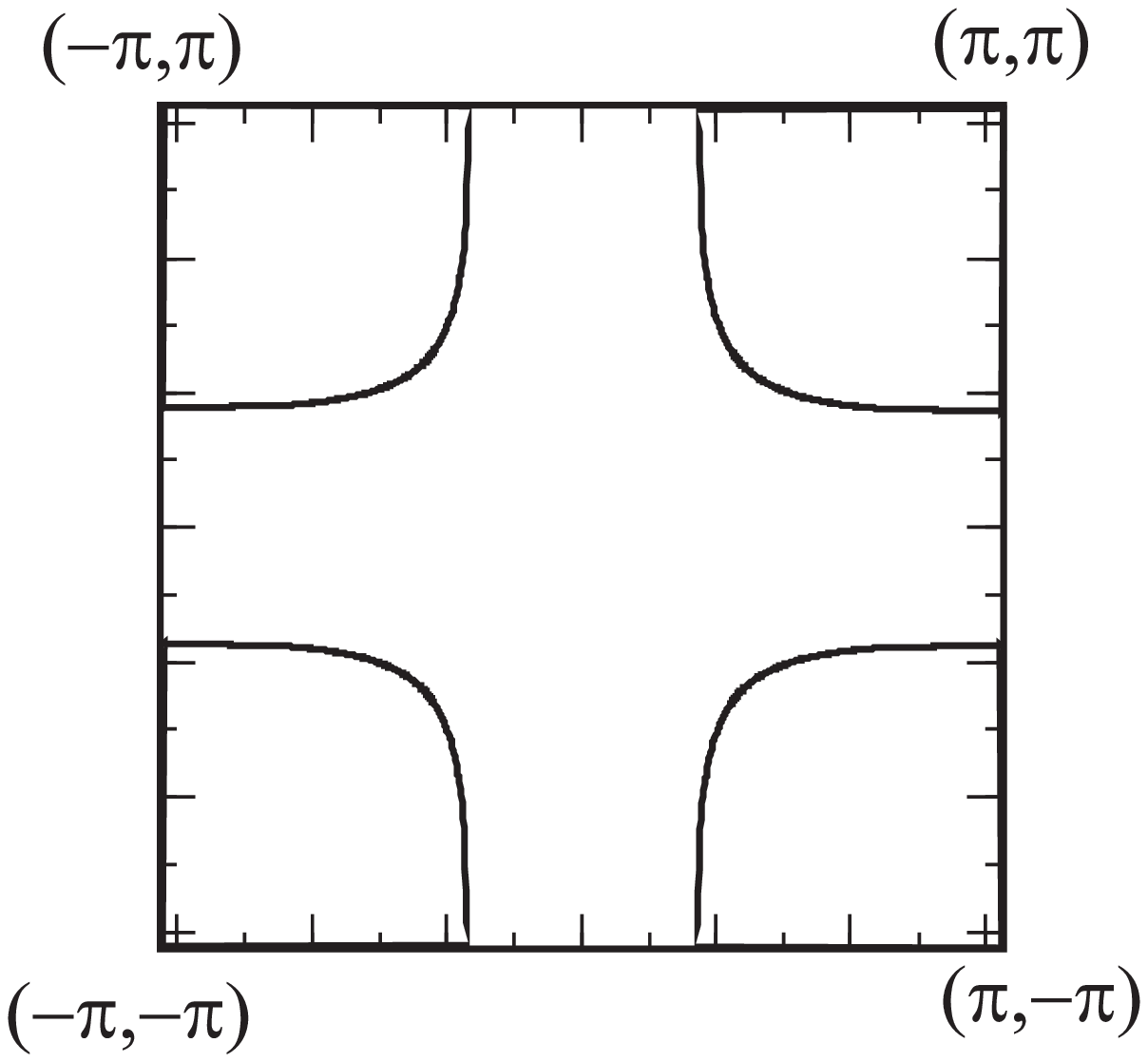}
\includegraphics[width=0.4\columnwidth]{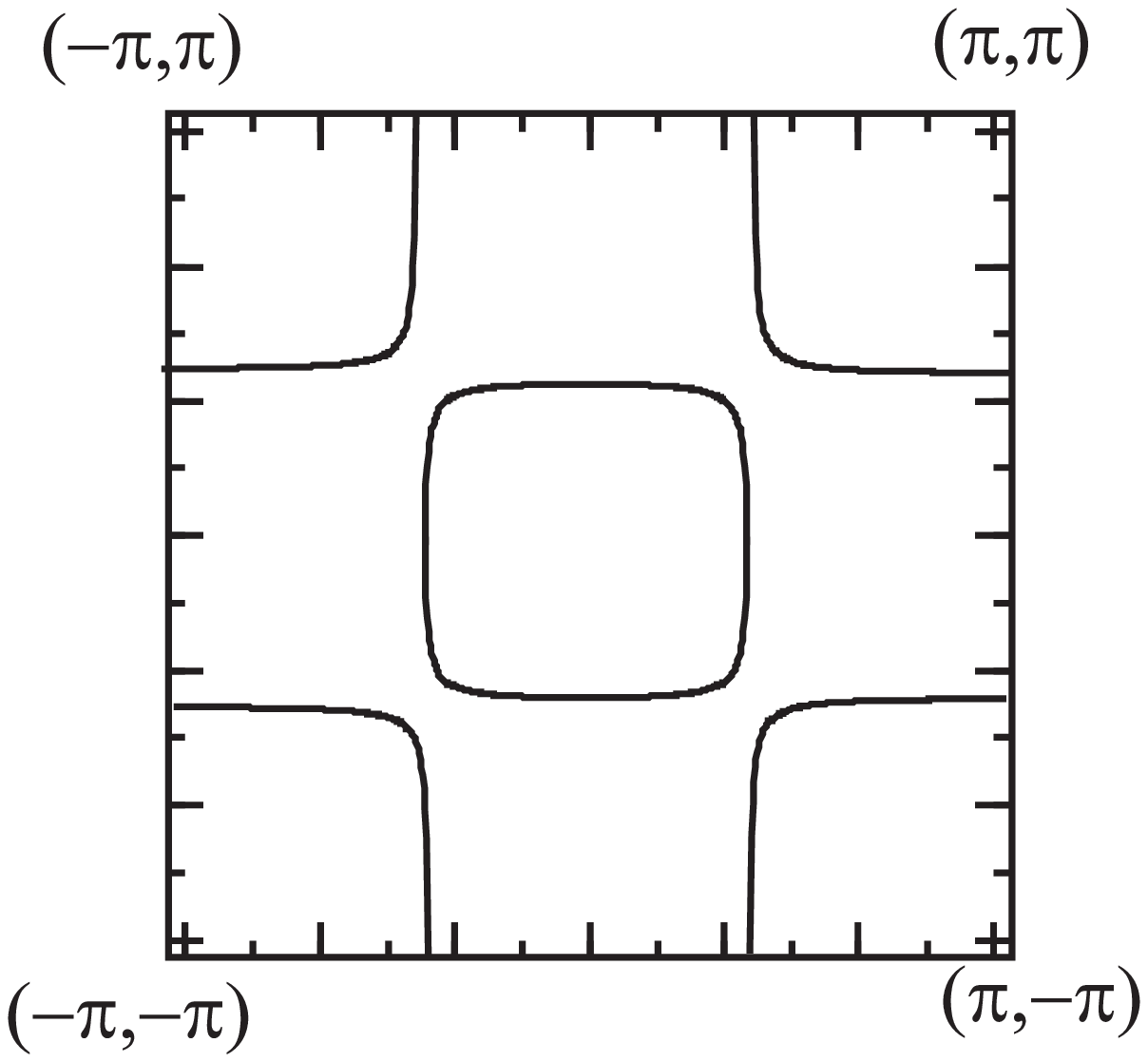} \\
\vspace{9pt}
  \hbox{\hspace{0.1\columnwidth} (c) \hspace{0.4\columnwidth} (d)}
  \vspace{9pt}
\caption{Evolution of the Fermi-surface with increasing the
electron concentration $n$ } \label{fig:FS}
\end{figure}

 Using
\begin{eqnarray}
&& a^{\dagger}_{1} = \sum_i e^{i\mathbf{p R}_i}
(\alpha_{\mathbf{p}}c^{\dagger}_{|x^2-y^2\rangle i}+
\beta_{\mathbf{p}}c^{\dagger}_{|z^2\rangle i}) \\ &&
a^{\dagger}_{2} = \sum_i e^{i\mathbf{p R}_i}
(-\beta_{\mathbf{p}}c^{\dagger}_{|x^2-y^2\rangle i}+
\alpha_{\mathbf{p}}c^{\dagger}_{|z^2\rangle i})
\label{eq:Bogolubov}
\end{eqnarray}
it is straightforward to calculate the band dispersion and total
energy of different phases (cf. \cite{vandenBrink99a,Efremov03}).
For example for the simplest case of A-type  antiferromagnetic
ordering one has the following bands:
\begin{eqnarray} \label{eq:specA}
\varepsilon_{1,2}(\mathbf{p})/t =&&  - \bigg[ \cos p_x + \cos p_y   \\
&&  \pm \left( \cos^2 p_x + \cos^2 p_y - \cos p_x \cos p_y
\right)^{1/2} \bigg],  \nonumber
\end{eqnarray}
and for the ferromagnetic case one has
\begin{eqnarray} \label{eq:specF}
\varepsilon_{1,2}(\mathbf{p})/t  = && - \bigg[ \cos p_x + \cos p_y
+ \cos p_z
\\  && \pm  \left( \cos^2 p_x + \cos^2 p_y + \cos^2 p_y -
 \cos p_x \cos
p_y
 \right.  \nonumber \\ &&  \left.  - \cos p_x \cos p_z- \cos p_y \cos p_z \right)^{1/2} \bigg]
\nonumber
\end{eqnarray}
It is instructive to look at the
form of the Fermi-surface for different concentration of
electrons. For the A-type magnetic
structure with increasing number of electrons $n$ in $e_g$-bands
($n=1-x$ in La$_{1-x}$Ca$_x$MnO$_3$) we first start to fill the band $\varepsilon_{1}$ in
(\ref{eq:specA}). It has predominantly
$|x^2-y^2\rangle$- character close to the
$\Gamma$-point (center of the Brillouin zone). The second band
$\varepsilon_{2}$ begins  to fill  from $n\approx 0.14$, creating
the second pocket of the Fermi-surface. The evolution of the
Fermi-surface is shown schematically in Fig. \ref{fig:FS}

\begin{figure}
\includegraphics[width=0.42\columnwidth]{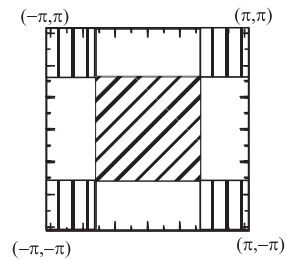}
\caption{Fermi surface in the limit $n \to 1$. The "electron" pocket is shown by the diagonal hatching;
"hole" -- by vertical}.
 \label{fig:cross}
\end{figure}

We see that by approaching  $n=1$ (the case of the undoped LaMnO$_3$) the shape of
the Fermi-surface pockets becomes more and more "flat". Although for
$n=1$ one can not speak about electrons and holes in strict sense, from the sequence of the filling ( Figs. 1 a-d) it is clear that one
should treat the inner pockets of the Fermi-surface as
electron, and the outer one - as hole-like. It is true at least for $n$  slightly less than one.

It is also clear that for $n=1$ there exist perfect nesting
between the bands 1 and 2: their Fermi-surfaces coincide with the shifting by the wavevector $Q=(\pi, \pi )$.
It can be seen directly from the Eq. \ref{eq:specA}.  For  $n=1$  one has
\begin{equation}\label{eq:nesting}
\varepsilon_{1} (\mathbf{p}+\mathbf{Q}) = -
\varepsilon_{2}(\mathbf{p})
\end{equation}
The spectrum (\ref{eq:specF}) for the ferromagnetic phase (3D - case)
obeys the same relation (\ref{eq:nesting}) with the wave vector $Q=(\pi,\pi,\pi)$.
 Hence  the full nesting
between the bands $1$ and $2$ take place also in the
ferromagnetic case, although  the Fermi surface is not flat anymore.

\section{Excitonic instability and pairing}

It is well known, that in the situation of nesting
the original homogenous metallic state of the system is unstable.
Depending on the dominant interactions
different types of ordering state may exist: those with singlet or
triplet, real or imaginary order parameters, etc.
\cite{Halperin}

In order to study the consequences of nesting for magnetically
ordered materials we assume the magnetic order of localized spins
of A-type. We want to note that due to the large Hund's coupling
there is no hopping from one ferromagnetic plane to another and
the spin indexes are fixed. But instead of the spin degeneracy,
which is lifted, we have here another degree of freedom - double
orbital degeneracy, which plays the role of spins in the
nondegenerate case. Thus for the repulsive electron-electron
interaction, which we assume, we would have, instead of the
formation of a spin density wave (SDW) state, the corresponding
formation of its orbital analog - orbital density wave (ODW). We
will show that this state is indeed realized in our case. The new
ground state in case of nesting  can be easily found in the
standard Green function technique (cf. for example
\cite{AGD,Keldysh}). We will consider the case of the A-phase. The
ferromagnetic case may be studied along the same lines.

Due to coupling between the two bands (\ref{eq:specA}) the
anomalous Green function arises. It is obvious that while normal
Green functions $G^{11}(\mathbf{p})$ and $G^{22}(\mathbf{p})$ are
proportional to $\delta_{p,  p'}$, the anomalous Green function
$G^{12}(\mathbf{p}, \mathbf{p'})$ is proportional to $\delta_{p+Q, p'}$. Further we will use notation
$G^{12}(\mathbf{p})$ for the sake of simplicity.
 In
general the equations for the Green functions can be written in
the following form:
 \ben ( i \omega_n - \eps_2(\mathbf{p})+\mu )
G^{22}(\mathbf{p}, \omega_n) = 1 + \Sigma^{12}(\mathbf{p},
\omega_n) G^{12}(\mathbf{p},\omega_n) \een
 \ben ( i \omega_n -
\eps_1(\mathbf{p})+\mu ) G^{12}(\mathbf{p}, \omega_n) =
\Sigma^{12}(\mathbf{p}, \omega_n)G^{22}(\mathbf{p},\omega_n) \een
where
 \beq
&&\Sigma^{12}(\mathbf{p},\omega_n) = \\&& - T \sum_{m} \int
\frac{d \mathbf{p} }{(2 \pi)^{3}} \Gamma^{12}(\mathbf{p},
\mathbf{p'},\omega_n -\omega_m) G^{12}(\mathbf{p}, \omega_m)
\nonumber
 \eeq
with $\Gamma^{12}(\mathbf{p}, \mathbf{p'},\omega_n -\omega_m)$
being  the repulsive electron-electron  interaction. Using $\eps_1
(\mathbf{p}) = - \eps_2(\mathbf{p}+ \mathbf{Q}) = \eps(\mathbf{p})
$ one can write:
\begin{eqnarray}\label{eq:KeldyshKopaev}
&& G^{22}(\mathbf{p},\omega_n) =  -  \frac{ i \omega_n+
\varepsilon({\mathbf{p}})}{\omega^{2}_n+
\varepsilon^2(\mathbf{p})+|\Sigma^{12}(\mathbf{p},\omega_n)|^2}
\\
&& G^{12}(\mathbf{p},\omega_n)=  -
\frac{\Sigma^{12}(\mathbf{p},\omega_n)}{\omega^{2}_n+
\varepsilon^2(\mathbf{p})+ |\Sigma^{12}(\mathbf{p},\omega_n)|^2}
\\ &&
\Sigma^{12}(\mathbf{p},\omega_n) =  -  T\sum \int \frac{d
\mathbf{p'}}{(2 \pi)^{3}} \Gamma^{12}(\mathbf{p},
\mathbf{p'},\omega_n -\omega_m)   \nonumber \\ && \times
\frac{\Sigma^{12}(\mathbf{p},\omega_m)}{\omega^{2}_m+
\varepsilon^2(\mathbf{p})+ |\Sigma^{12}(\mathbf{p},\omega_m)|^2}
\end{eqnarray}
Assuming  $\Gamma^{12}(\mathbf{p},\mathbf{p}', \omega_n -
\omega_m)$ being independent on frequency one gets that
$\Sigma^{12}(\mathbf{p}, \mathbf{p}', \omega_n,\omega_m)$ is also
frequency-independent. The quantity

\be \label{eq:SpectrumODW} E  = \pm
\sqrt{\varepsilon(\mathbf{p})^2+ \Delta^2(\mathbf{p})} \ee gives
the spectrum of elementary excitations with the gap
$|\Delta(\mathbf{p})| = |\Sigma^{(12)}(\mathbf{p})|$.
 The gap can
be determined from the following equation
 \be
\Delta(\mathbf{p}) = \int \frac{d \mathbf{p}'}{(2 \pi)^{2}}
\Gamma^{12}(\mathbf{p}, \mathbf{p}')
 \frac{\Delta(\mathbf{p}')}{2 E(\mathbf{p}')} \tanh \frac{E(\mathbf{p})}{2 T },
 \ee
which in the presence of the nesting always has nontrivial
solution. For the simple on-site interaction (Hubbard repulsion $U
n_{i \alpha \uparrow} n_{i \beta \uparrow}$ with $\alpha \neq
\beta$) one gets in the weak-coupling case $U \ll W$ the usual
solution with $\Delta(T=0) \sim \varepsilon_F \exp(-W/U)$ with $W$
being bandwidth. One can also show that the results do not
crucially depend on the choice of Hubbard interaction, and that
they remain qualitatively the same for the electron-electron
repulsion of the screened Coulomb type with the screening length
larger than the intersite distance.

 As we  see from Eq.(\ref{eq:SpectrumODW}), the
resulting spectrum has a gap and consequently this state is
insulating, in agreement with the insulating nature of LaMnO$_3$.
It is also clear that the pairing with the wave vector
$Q=(\pi,\pi)$ leads to the formation of the checkerboard-type
superstructure in the xy-plane.

Using the inverse transformation of (\ref{eq:Bogolubov}) one can
find the alternation of the occupation numbers of different
orbital states due to  interband excitonic pairing (appearance of
the anomalous average $\delta n^{(i)}_{|x^2-y^2\rangle |z^2\rangle
} = \langle c^{\dagger}_{|x^2-y^2\rangle i }c_{|z^2\rangle i }
\rangle$) as:
\begin{eqnarray}
&& \delta n^{(i)}_{|x^2-y^2\rangle  }  = \delta
n^{(i)}_{|z^2\rangle   }  =0
\\ &&
\delta n^{(i)}_{|x^2-y^2\rangle |z^2\rangle  }  = (-)^{i} D,
\end{eqnarray}
where $D= G^{12}(x,x)$ and proportional to the energy gap.

The obtained excitonic insulator state corresponds to the
formation of an orbitally-ordered state in which in each plane there
appears two orbital sublattices of the orbitals $| \phi \rangle =
\cos (\phi/2) |z^2 \rangle \pm \sin (\phi/2) | x^2-y^2\rangle $
with the angle
\begin{eqnarray}
&& \tan \phi/2  = \\
&&
\frac{(n_{|z^2\rangle}-n_{|x^2-y^2\rangle}))-\sqrt{(n_{|z^2\rangle}-n_{|x^2-y^2\rangle})^2+4
D^2}}{2 D}, \nonumber
\end{eqnarray}
where the diagonal averages $n_{z^2}$ and $n_{x^2-y^2}$ for $n=1$
are slightly different: $n_{z^2} \approx 0.45$, $n_{x^2-y^2}
\approx 0.55$ due to specific form of occupied bands. For equal
occupation $n_{z^2} = n_{x^2-y^2} = 0.5$ the ordering would
correspond to the angle $\phi = \pi/2$, i.e. to the equal mixture
of the basis orbitals ($|\pi/2> = 1/\sqrt{2}(|z^2 \rangle \pm
|x^2-y^2 \rangle)$).

Experimentally it was found that the corresponding orbitals in
LaMnO$_3$ are $|\phi \sim \pm 100 ^{\circ} \rangle$; this solution
in our analysis would be obtained for intermediate Hubbard
interaction $W/U \sim 3$. We would like also to note that in the
standard treatment of orbital ordering in the picture of localized
electrons one obtains close to $T_c$ similar  $|\pm \pi/2 \rangle$
orbital sublattices \cite{Feiner}).

Turning to the ordering in the third direction, for the real
A-type magnetic structure we see that in the accepted
approximation (no hopping and no dispersion in $z$-direction) the
system "does not know" how to order. If we include certain weak
hopping in the $z$-direction which could appear at finite
temperatures and, even at $T=0$, due to quantum corrections to the
leading terms considered here \cite{Kagan}, these hopping would
lead to the appearance of the main instability at the wavevector
$Q =(\pi, \pi, \pi)$, i.e. it would give orbital alternation also
in the $z$-directions. The same type of ordering appears also in
the ferromagnetic case. Apparently the experimentally observed
ordering, which in LaMnO$_3$ is that with the ``in phase'' orbital
ordering in the $z$-direction, $Q =(\pi, \pi, 0)$, is caused by
some other factors, e.g. an interplay of GdFeO$_3$ distortion
(tilting of MnO$_6$-octahedra) and Jahn-Teller deformations
\cite{Mizokawa}.

Referring to the theoretical situation, we should note that in
contrast to the usually considered excitonic insulators, for the
A-type magnetic structure there exist not only an interband
nesting, but for n=1 also each pockets has nesting within itself,
with its own vectors $Q_{x,y} = (\pi, 0),(0, \pi)$ (this was also
noticed by Hotta \cite{Hotta}, who, however, did not consider the
interband instability). In principle  these two instabilities may
coexist and influence one another.

 One can show that the inclusion of both these factors just strengthens
 the excitonic instability and increases the resulting energy gap,
 without changing qualitative features of the solution. Besides,
 keeping in mind the sequence of band
fillings shown in Figs.1 a-d, we expect that in the realistic
situation the interband pairing would dominate and the results
presented above would be valid. Besides, intraband nesting exist
for $n=1$ for the A-type structure, but it is absent for the
ferromagnetic case.

\section{Conclusions}

Thus we see that one can explain, at least qualitatively, the
appearance of an orbital ordering and opening of the gap in the
electron energy spectrum of undoped LaMnO$_3$ and similar systems
proceeding from the band picture. In this approach  orbital
ordering appears due to an excitonic instability, caused by the
nesting between pocket of two $e_g$-bands. The resulting state has
a correct type and symmetry of the orbital ordering. Thus one can
interprets the ground state of LaMnO$_3$ as arising due to this
excitonic instability.

Of course, as already mentioned in the introduction, we do not
claim that our treatment gives the only correct picture of
LaMnO$_3$; there are many indications that the electron
correlations in it are relatively strong, so that it is usually
rightly treated as a Mott insulator. However the truth is often
``in between'', the system may show the features of the both the
limiting pictures. In any case it is gratifying that one can
obtain a quite  reasonable description of this system proceeding
from the band point of view. This result, in certain sense,
completes similar investigations carried out in
\cite{vandenBrink99a,vandenBrink99b,Efremov03}. In these papers it
was shown that the band approach works quite well for description
of orbital structure of overdoped, half-doped and
less-than-half-doped manganites. It is even somewhat surprising
that using the same simple approach one can explain even the
properties of an undoped LaMnO$_3$, which until now was always
considered as a bona fide Mott insulator. One may hope that this
novel point of view may prove useful both for better understanding
of the theoretical situation and for the interpretation of certain
experiments.

We are grateful to  P. Fulde and
M. Mostovoy  for useful discussion. This work was supported by the Netherland Foundation for the Fundamental Research of Matter (FOM)  and by the Deutsche Forschungsgemeinschaft via SFB 608.


\begin{thebibliography}{10}
\bibitem{orbitalnote}  We denote the $3z^2-r^2$ orbital by $z^2$, $3y^2-r^2$ by $y^2$ etc.
\end{thebibliography}
\end{document}